# Deterministic generation and partial retrieval of a spin-wave excitation in an atomic ensemble

SHENGZHI WANG,[1,2] ZHONGXIAO XU,[1,2,3] DENGXIN WANG,[1,2] YAFEI WEN,[1,2] MINJIE WANG,[1,2] PAI ZHOU,[1,2] LIANG YUAN,[1,2] SHUJING LI,[1,2] AND HAI WANG[1,2,*]

[1]*The State Key Laboratory of Quantum Optics and Quantum Optics Devices, Institute of Opto-Electronics, Shanxi University, Taiyuan 030006, China*

[2]*Collaborative Innovation Center of Extreme Optics, Shanxi University, Taiyuan 030006, China*

[3] *xuzhongxiao@sxu.edu.cn*

\* *wanghai@sxu.edu.cn*

**Abstract:** In this paper, we report a generation of a spin-wave excitation (SWE) with a near-unity ( $0.996 \pm 0.003$ ) probability in a given time (~730 $\mu s$ ). Such deterministic generation relies on a feedback scheme with a millisecond quantum memory. The millisecond memory is achieved by maximizing the wavelength of the spin wave and storing the SWE as the magnetic-field-insensitive transition. We then demonstrate partial retrievals of the spin wave by applying a first read pulse whose area is smaller than the value of $\pi$ . The remained SWE is fully retrieved by a second pulse. Anti-correlation function between the detections in the first and second readouts has been measured, which shows that the partial-retrieval operation on the SWE is in the quantum regime. The presented experiment represents an important step towards the realization of the improved DLCZ quantum repeater protocol proposed in Phys. Rev. A 77, 062301 (2008).

## 1. Introduction

The distribution of entanglement states over long distances is crucial for quantum communication and large-scale quantum networks [1, 2]. However, due to unavoidable transmission losses in quantum channels and the no-cloning theorem for quantum states, such distribution task is still a challenge. One possible solution is the use of quantum repeaters (QRs) [3]. The basic ideal of the QR protocol is dividing the long distance L into N shorter elementary links. Entanglement generation attempt is independently performed in each link. Quantum memories are required for storing the created entanglement in a link until entanglement has been established in the adjacent links. By entanglement swapping between the adjacent links, entanglement will be ultimately extended to the full distance [3].

The Duan-Lukin-Cirac-Zoller (DLCZ) protocol [4] , which is using atomic-ensemble-based quantum memory, is attractive because it is relatively simple [2].The

DLCZ scheme relies on spontaneous Raman scattering [2, 4], which can probabilistically create a pair of correlated excitations, namely, one spin-wave excitation (SWE) and a single photon. The spin-wave excitation can be stored in an atomic ensemble and the photon can be sent to a center station between two remote atomic ensembles. Based on the detection of a photon which could come from either of the two remote atomic ensembles, the entanglement between the remote memories can be established. Along with the DLCZ scheme, many experiments have demonstrated the generation of the pair of correlated excitations via spontaneous Raman scattering (SRS) [5-14]. The lifetime and retrieve efficiency of the quantum memories have been significantly improved. For example, the retrieval efficiency and lifetime of storing a single spin wave in laser-cooled $Rb^{87}$ atoms could up to 76% and 3ms [10]. However, the DLCZ protocol has some practical drawbacks. On the one hand, the entanglement generation via single-photon detections requires long-distance phase stability, which is a technique challenge [2]. On the other hand, in order to suppress multi-excitations, the probabilities of preparing the spin-wave-photon quantum correlation or entanglement have to keep very low [2, 15], which lead to a very low quantum repeater rate [15]. To overcome these drawbacks, several improved DLCZ QR protocols have been proposed [15-19]. The proposed protocol in [15], N. Sangouard et al shows that the local creation of high-fidelity entangled pairs of spin-wave excitations (SWEs) in combination with the use of two-photon detections for remote entanglement generation, holds promise to implement much robust and efficient quantum repeaters. In this improved protocol, the local entanglement creation relies on partial read-out operations on four DLCZ-like quantum memories. The procedure of this creation may be explained in the following. Via the detection of a single Stokes photon emitting from one atomic ensemble, one may herald the storage of one spin-wave excitation in the atomic ensemble. After the four ensembles are charged, the spin-wave excitations are simultaneously partially readout, creating a probability amplitude to emit an anti-Stokes photon. Based on a coincident detection of two anti-Stokes photons at a polarizer beam splitter, the atomic ensembles will be projected into an entangled state [15]. The key requirement in such local entanglement creation is that the SWE memory in each ensemble can be deterministically created within a given time and then be partially readout at a predetermined time. The deterministic creation of one SWE has been demonstrated in several previous experiments [12-14]. Limiting to the storage lifetime of $20-30\mu s$, the creation probabilities of the SWE are in a range of ~10-30% in these experiments, which still remain an improvement. In a recent experiment, Li, J. et al. demonstrated deterministic generation of single SWEs via Rydberg blockade in atoms [20]. However, due to experimental imperfections, such as spatial intensity inhomogeneity of the manipulation laser beams and fluctuation of atom numbers, the preparation probability for single excitation is about 55%. The partially read out of single Rydber SWEs [21] and ground-state SWEs [22] have been experimentally demonstrated, respectively. However, the autocorrelation function between two partially readouts of a single SWE has not been experimentally demonstrated yet.

Here, we achieve a millisecond quantum memory by maximizing the spin-wave wavelength and selecting a magnetic-field-insensitive transition to store the spin wave (SW) [7]. Based on the long-lived quantum storage and detection-based feedback scheme, we generate one spin-wave excitation (SWE) with a near-unity probability within a given time of ~730 $\mu s$. By applying a read pulse whose area is less than $\pi$, we demonstrate partial read out of the SWE. The remained SWE is fully retrieved by a second read pulse. The anti-correlation function between the first and second readouts is measured when the retrieval efficiencies for two retrievals are the same. The measurement result shows that the readout operations on the SWE are in single-quanta regime.

## 2. Experimental setup and analysis

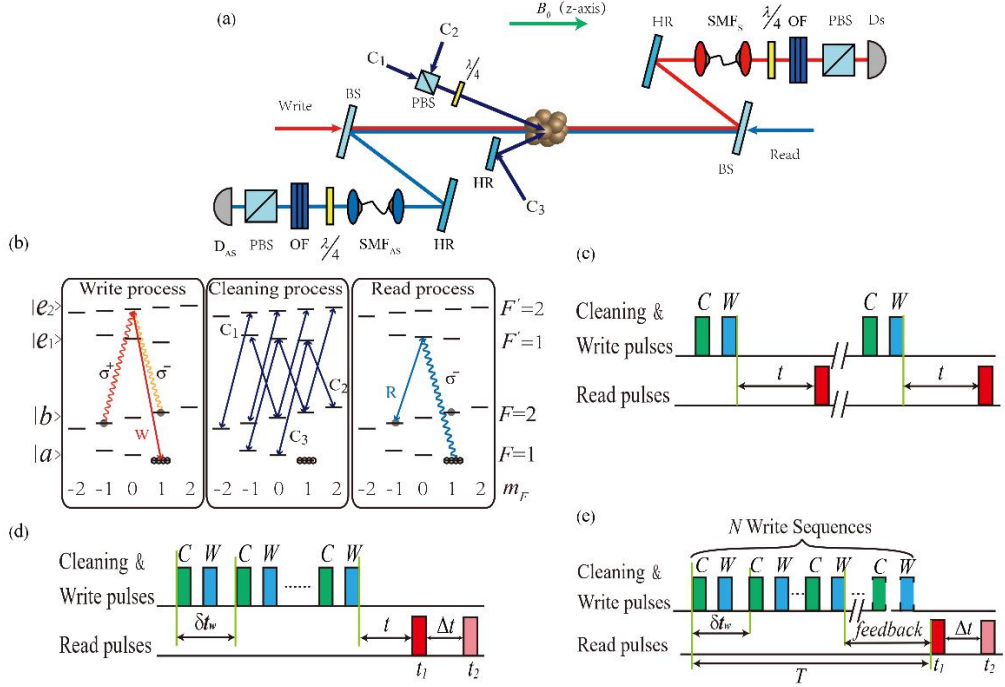

Fig. 1. (a) Illustration of the experimental setup. HR: high reflection mirror; PBS: polarization beam splitter; SMF: single-mode fiber; $D_S$ ($D_{AS}$): Stokes (anti-Stokes) single-photon detector; BS: beam splitter; OF: optical filters, (b) Relevant atomic levels, (c) Time sequence for measuring the correlation function and retrieval efficiency, (d) Time sequence for measuring the anti-correlation parameter, (e) Time sequence of the measurement-based feedback protocol. C: cleaning laser, which includes pumping laser beams C1, C2 and C3. W: write laser.

The experimental setup is shown in Fig. 1(a). The atomic ensemble is a cloud of cold $^{87}$Rb atoms, whose relevant atomic levels are shown in Fig. 1(b), where $|a\rangle = |5^2 S_{1/2}, F=1\rangle$, $|b\rangle = |5^2 S_{1/2}, F=2\rangle$, $|e_1\rangle = |5^2 P_{1/2}, F'=1\rangle$ and $|e_2\rangle = |5^2 P_{1/2}, F'=2\rangle$. Each level includes Zeeman sublevels, for example, $|a\rangle$ is rewritten as $|a, m_F\rangle$, where the magnetic quantum number $m_F = \pm 1, 0$. We apply a moderate magnetic field $B_0 = 13.5G$ along z-direction [as shown in Fig. 1(a)] to define the quantum axis and lift the degeneracy of Zeeman sublevels in the levels. Such that the Zeeman sublevels are split, which enable us to efficiently prepare the atoms into the level $|a, m_F = 1\rangle$ by optical pumping. The writing pulses with 10MHz red-detuned to $|a\rangle \rightarrow |e_2\rangle$ transition are applied on to the atoms, which induce the Raman transition $|a, m_F = 1\rangle \rightarrow |b, m_F = -1\rangle$ or $|a, m_F = 1\rangle \rightarrow |b, m_F = 1\rangle$ [Fig. 1(b)] via $|e_2, m_F = 0\rangle$ level, which emits a $\sigma^+$-polarized or $\sigma^-$-polarized Stokes photon and creates magnetic-field-insensitive SW associated with the coherence $|a, m_F = 1\rangle \rightarrow |b, m_F = -1\rangle$ or magnetic-field-sensitive SW associated with the coherence $|a, m_F = 1\rangle \rightarrow |b, m_F = 1\rangle$. The Stokes photons are collected by a single-mode fiber $SMF_S$. Thus they have the wave-vector $k_S$. The wave-vector of the SW mode is given by $k_{SWE} = k_W - k_S$, where, $k_W$ is the wave-vector of the write pulse. After the single-mode fiber $SMF_S$, we transform the $\sigma^+$-polarized or $\sigma^-$-polarized state of the Stokes photons into the horizontally- (*H-*) or vertically- (*V-*) polarized state by placing a λ/4 plate in their paths [see Fig. 1(a)]. Then, the Stokes photons go through optical filters which are used for greatly attenuating the write laser field [23] and then direct to a single-photon detector labeled by $D_S$. Before the detector $D_S$, we use a polarization beam splitter (PBS) to block the *V*-polarized Stokes photons into the detector $D_S$. Thus, only the *H*-polarized photons may arrive in the detector $D_S$. In this case, the detection of a photon at the detector $D_S$ heralds the storage of one magnetic-field-insensitive SWE. For suppressing the dephasing effect resulting from atomic motions, we make the Stokes field and writing light beams collinearly go through the cold-atom cloud along the *z*-direction [see Fig. 1(a)]. The uses of magnetic-field-insensitive SW and the collinear propagations promise us to achieve a long-lived SWE storage. The SWE can be efficiently converted into an anti-Stokes photon by applying a reading laser pulse with its frequency being tuned to the transition $|b, m_F = -1\rangle \leftrightarrow |e_1, m_F = 0\rangle$. The reading laser pulses, labeled by *R*, propagate along the *z*-axis, which has the wave-vector of $k_R \approx -k_W$. According to the phase matching condition, the wave-vector of the retrieved anti-Stokes photon is $k_{AS} = k_W - k_S + k_R \approx -k_S$, means that the propagation direction of the anti-Stokes photon is also along the *z*-axis. We use a single-mode fiber $SMF_{AS}$ to collect the anti-Stokes photons. After the $SMF_{AS}$, the $\sigma^+$ or $\sigma^-$-polarized anti-Stokes photons are transformed into *H*- or *V*- polarized photons by a λ/4 plate [see Fig. 1(a)]. Then, the anti-Stokes photons go through optical filters and then are sent to a single-photon detector, labeled by $D_{AS}$ in Fig. 1(a).

## 3. Experimental results

For showing the lifetime of the storage of the SWE as well as the quantum correlation between the Stokes photon and the SWE in our memory system, we measure the retrieve efficiency and the cross-correlation function between Stokes and anti-Stokes fields as the function of the storage time *t*. The time sequence for the measurement is shown in Fig. 1(c), where, the experimental run, which includes n trails is performed after the atoms are loaded to a magneto-optical trap (MOT). Each trail contains a cleaning pulse, a write pulse and a read pulse. The cleaning pulse is used to prepare (or pump) the atoms to (or back to) the state $|a, m_F = 1\rangle$. The cleaning laser includes three pumping laser beams, which are labeled by C1, C2 and C3 in the Fig. 1(a). The $\sigma^+$-polarized C1 laser beam and $\sigma^-$-polarized C2 laser beam, whose frequencies are tuned on the $|b\rangle \leftrightarrow |e_2\rangle$ and $|b\rangle \leftrightarrow |e_1\rangle$ transitions, respectively, are combined on a polarization-beam-splitter (PBS) and then go through the atoms at an angle of about 3° relative to the *z*-axis. The $\sigma^+$-polarized C3 laser beam, whose frequency is tuned to the $|a\rangle \leftrightarrow |e_1\rangle$ transition, goes the atoms at an angle of about $-3°$ relative to the *z*-axis. The three laser beams C1, C2 and C3 overlap at the center of the atoms.

In the measurement, the stored SWE is fully retrieved and converted into an anti-Stokes photon by a read pulse with a duration of $\delta t = 50 ns$ and a power of $P_0 = 300 \mu W$ after a storage time *t*. The read pulse area for such a case is $\pi$. The cross-correlation function is defined as $g_{AS,S}^{(2)} = p_{AS,S} / p_{AS} p_S$, where, $p_S$ ($p_{AS}$) denotes the probability of detecting one Stokes (anti-Stokes) photon and $p_{AS,S}$ is the coincident probability between the Stokes and anti-Stokes photons. Considering the background noise in each channel, we have $p_S = \chi\eta_S + B\eta_S$, $p_{AS} = \chi\gamma\eta_{AS} + C\eta_{AS}$ and $p_{AS,S} = \chi\gamma\eta_S\eta_{AS} + p_S p_{AS}$, where, $\chi$ is the excitation probability per write pulse, $B$ ($C$) denotes background noise in the Stokes (anti-Stokes) channel, which is far less than $\chi$ in the presented system, $\eta_{AS}$ ($\eta_S$) is the overall detection efficiencies in the anti-Stokes (Stokes) channel, whose values are $\eta_{AS} = \eta_S \approx 0.2$ in the presented experiment, $\gamma$ is the retrieve efficiency and is measured according to $\gamma \approx p_{AS,S}/(\eta_{AS} \cdot p_S)$. Considering the decoherence of the SW, the retrieve efficiency $\gamma$ will decay as the storage time and can be described as [23]:

$$\gamma(t) = \gamma_0 e^{-t/\tau_0}, \tag{1}$$

where, $\gamma_0$ is zero-delay retrieval efficiency and $\tau_0$ is the storage life-time. When the storage time is very long, $\gamma$ will go to zero and then $p_{AS}$ approach $C\eta_{AS}$. Based on the expressions of $p_S$, $p_{AS}$, $p_{AS,S}$ and the cross-correlation function $g^{(2)}_{AS,S}$, the cross-correlation function is rewritten as

$$g^{(2)}_{AS,S}(t) = 1 + \frac{\gamma(t)}{(\chi+B)\gamma(t)+D}, \tag{2}$$

where, $D = C(1+B/\chi)$. Due to $B \ll \chi$, we have $D \approx C$. For ideal case of $B \approx C \approx 0$, the equation (2) may be rewritten as $g^{(2)}_{AS,S}(t) = 1 + 1/\chi$.

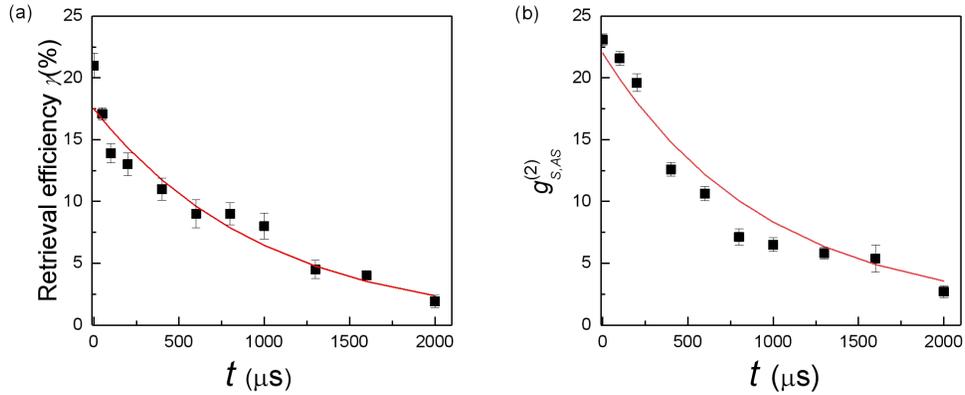

Fig. 2. Measured revival efficiency $\gamma$ (a) and cross-correlation function $g^{(2)}_{AS,S}$ (b) as a function of the storage time $t$ for $\chi = 3\%$, $P_0 = 300\mu W$ and $\delta t = 50 ns$. Error bars represent 1 standard deviation.

We then measure the revival efficiency $\gamma$ and the cross correlation $g^{(2)}_{AS,S}$ as a function of the storage time $t$. The black square dots in Figs. 2(a) and 2(b) are the measured $\gamma$ and $g^{(2)}_{AS,S}$ as a function of the storage time $t$ for the excitation probability $\chi = 3\%$, respectively. The solid red curves in Figs. 2(a) and 2(b) are the fittings based on the equations (1) and (2), respectively, which yields the storage life-time $\tau_0 = 900\mu s$ and zero-delay retrieval efficiency $\gamma_0 = 17.5\%$. From the Fig. 2(b), we can see that $g^{(2)}_{AS,S}$ is well above 2 at $t = 2ms$, meaning that the SWE and the Stokes photon are non-classically correlated [2] at this time.

The detection of a Stokes photon at $D_S$ will herald the creation of one magnetic-field-insensitive SWE, which is denoted by $s_0^\dagger |0\rangle$, where $s_0^\dagger$ ($|0\rangle$) refers to the creation operator of the SWE (vacuum state). For performing the scheme in [15], N. Sangouard et al, the SWE is required to be partially readout, which may be done by using the read pulse whose area is smaller than $\pi$. We may demonstrate the partial readout by using the read pulse whose duration is kept unchanged ($\delta t = 50 ns$) and power is set to be less the fully-read-out value $P_0 = 300 \mu W$. Assuming that such a read pulse, denoted by $R_1$ (first read pulse), is applied onto the atoms after the storage time $t_1$, it will convert the SWE $s_0^\dagger |0\rangle$ into an anti-Stokes photon in the time bin $AS_1$, with a partial retrieval efficiency $|\varsigma|^2 \gamma_1$, where, $|\varsigma|^2 \leq 1$ is the partial read-out weight and $\gamma_1 = \gamma(t_1) = \gamma_0 e^{-t_1/\tau_0}$ is the retrieval efficiency. Such a process can be described as:

$$s_0^\dagger |0\rangle \xrightarrow{R_1} \left(\varsigma \sqrt{\gamma_1} b_{AS_1}^\dagger + \xi s_L^\dagger \right)|0\rangle, \qquad (3)$$

where $b_{AS_1}^\dagger$ is the creation operator of the anti-Stokes photon $AS_1$, $s_L^\dagger$ is the creation operator of the remained SW, $|\varsigma| = \sqrt{\gamma_{p_1}/\gamma_1}$, $\gamma_{p_1}$ is the retrieval efficiency when the SWE is partially read by $R_1$, which is measured according to the expression $\gamma_{p_1} \approx p_{AS_1,S}/(\eta_{AS} \cdot p_S)$, where, $p_{AS_1,S}$ is the coincident probability between the Stokes photon and the anti-Stokes photon $AS_1$, $\xi$ denote the probability amplitude of the remained SWE. The remained SW is fully converted into an anti-Stokes photon in the time bin $AS_2$ by applying the second read pulse $R_2$ with a power being equal to the fully-read-out power $P_0$ at time $t_2$, which corresponds to the process

$$\xi s_L^\dagger |0\rangle \xrightarrow{R_2} \xi \sqrt{\gamma_2} b_{AS_2}^\dagger |0\rangle = \sqrt{\gamma_{p_2}} b_{AS_2}^\dagger |0\rangle, \qquad (4)$$

where $b_{AS_2}^\dagger$ is the creation operator of the anti-Stokes photon $AS_2$, $\gamma_2 = \gamma(t_2)$ is the retrieval efficiency when the SWE is fully retrieved at $t_2$, $\gamma_{p_2}$ is the retrieval efficiency when the remained SWE is read out at $t_2$, which is measured according to $\gamma_{p_2} \approx p_{AS_2,S}/(\eta_{AS} \cdot p_S)$, here, $p_{AS_2,S}$ is the coincident probability between the Stokes photon $S$ and the anti-Stokes photon $AS_2$. According to the Eq.(4), we have $\xi = \sqrt{\gamma_{p_2}/\gamma_2}$. Assuming that the read time $t_1 \approx t_2 \approx 0$, we have $\gamma_1 = \gamma_2 = \gamma_0$. In this case, the total conversion process can be described as

$$s_0^\dagger |0\rangle \xrightarrow{R_1+R_2} \sqrt{\gamma_0} \left(\varsigma b_{AS_1}^\dagger + \xi b_{AS_2}^\dagger \right)|0\rangle, \qquad (5)$$

which shows that the SWE $s_0^\dagger |0\rangle$ may be converted into an anti-Stokes photon with a relative probability $|\varsigma|^2$ or $|\xi|^2$ in the time bin $AS_1$ (first readout) or the time bin $AS_2$ (second readout). Assuming that the loss of the SWE in the first readout is small and can be neglected, we have

$$|\varsigma|^2 + |\xi|^2 = \frac{\gamma_{p_1}}{\gamma_0} + \frac{\gamma_{p_2}}{\gamma_0} \approx 1. \qquad (6)$$

For demonstrating the above-described partial conversion, we experimentally measure the relative probability $|\varsigma|^2$ ($|\xi|^2$) as a function of the $R_1$-pulse power $P_1$ for $\delta t = 50 ns$. The black square dots (red circle dots) in Fig. 3 are the measured results of $|\varsigma|^2$ ($|\xi|^2$). From this figure, one may see that the probability $|\varsigma|^2$ ($|\xi|^2$) of the anti-Stokes photon in the first (second) readout is linearly increase (decrease) as the increase in the $P_1$. When $R_1$-pulse power $P_1 \approx 120 \mu W$ and its duration $\delta t = 50 ns$, we find that $|\varsigma|^2 \approx |\xi|^2 \approx 50\%$. The read pulse area for this case corresponds to $\pi/2$. The dashed line is the measured $|\varsigma|^2 + |\xi|^2$ as a function of $P_1$, which shows that $|\varsigma|^2 + |\xi|^2 \to 1$, satisfying the expectation in Eq. (6).

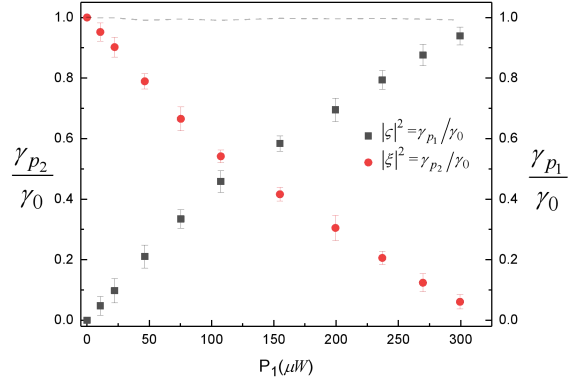

Fig. 3. Measured relative probabilities $|\varsigma|^2 = \gamma_{p1}/\gamma_0$ and $|\xi|^2 = \gamma_{p2}/\gamma_0$ in the first and second readouts as the function of the $R_1$-pulse power $P_1$ for $\delta t_R = 50 ns$. Error bars represent 1 standard deviation.

To demonstrate that our partial readout operate in the single-quanta regime, we measure the second order correlation functions between the detections at the time $t_1$ and $t_2$, which is defined as

$$\alpha = \frac{p_{S,AS_1,AS_2}}{p_{S,AS_1} p_{S,AS_2}}, \quad (7)$$

where, $p_{S,AS_1}$ ($p_{S,AS_2}$) refers to the probability of detecting an anti-Stokes photon in the time bin $AS_1$ ($AS_2$) conditional on a detection event of the Stokes photon and $p_{S,AS_1,AS_2}$ refers to the conditional coincidence probability between the detection events in the time bin $AS_1$ and $AS_2$. The above second order correlation function is the anti-correlation parameter of anti-Stokes photons, whose value $\alpha = 0$ corresponds to an ideal single photon and $\alpha = 1$ corresponds to classical light. The time sequence for the measurement of $\alpha$ is shown in Fig. 1(d), the write sequences, each of which contains a cleaning pulse and a write pulse, are applied to the atoms to generate the SWE. As soon as the detector $D_S$ detect a photon, the storage one SWE is heralded and the followed write sequences are terminal. After a storage time of $t$, i.e., at the time $t_1 = t$, the read pulse $R_1$ with an area of $\pi/2$ is applied to partially retrieve the SWE. At the time of $t_2 = \Delta t + t$ with $\Delta t = 1\mu s$, the read pulse $R_2$ is applied to fully retrieve the remained SW. The measurement results of $\alpha$ as a function of the storage time $t$ for $\Delta t = 1\mu s$ and $\chi = 3\%$ are shown in Fig. 4. From the Fig. 4, we can see that the measured $\alpha$ values are well below the classical bound of $\alpha = 1$ when $t \leq 1.3 ms$, which implies that the quantum nature of the SWE may be conserved for ~1.3ms.

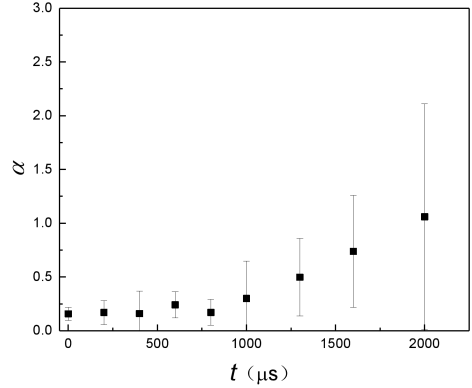

Fig. 4. The anti-correlation parameter $\alpha$ as a function of the storage time $t$ for $\Delta t = t_2 - t_1 = 1\mu s$. Error bars represent 1 standard deviation.

The probability of generating one SWE per write-pulse is equal to that of detecting a Stokes photon, which is $p_S = \chi \eta_S \approx 0.006$ in the above measurement. For the proposed scheme [15], the SW excitation is required to be able to be deterministically generated in a given time. To do this task, one may apply the measurement-based feedback protocol [13, 14]. The time sequence for the feedback protocol is shown in Fig. 1(e), where, a series of write sequences, each of which has a period of $\delta t_w = 1.6 \mu s$, applied to the atoms. Each write sequence contains a cleaning pulse and a write pulse. Once a Stokes photon is detected by $D_S$ at a write sequence, for example, at $j$-th write sequence, a field programmable gate array (FPGA), which is used for registering the detection event, will send out a feedback signal and then the further write sequence is stopped. At the predetermined time $T$ [see Fig. 1(e)], the read pulse $R_1$ is applied into the atoms to partially read out the SW. The time interval $T$ is limited to the memory lifetime $\tau$, which gives the maximum number of the trials $N = T / \delta t_w$. Thus the long-lived storage promises a large sequence number $N$. The measurement-based feedback protocol increases the detection probability of a Stokes photon from $p_S$ to $P_S = \sum_{i=0}^{N-1} p_S (1-p_S)^i$. The black open circle in Fig. 5 give the measured probability $p_S$ (i.e., the generation probability of one SWE) as a function of $N$ for $\chi = 3\%$. The solid red curve is the fit based on the function $P_S = \sum_{i=0}^{N-1} p_S (1-p_S)^i$ with $p_S = \chi \eta_S \approx 0.006$, where $\eta_S \approx 0.2$. From the Fig. 5, one can see that with the feedback-based protocol, the generation probability of one SWE approaches unity ($0.996 \pm 0.003$) for $N = 450$, $T = 730 \mu s$.

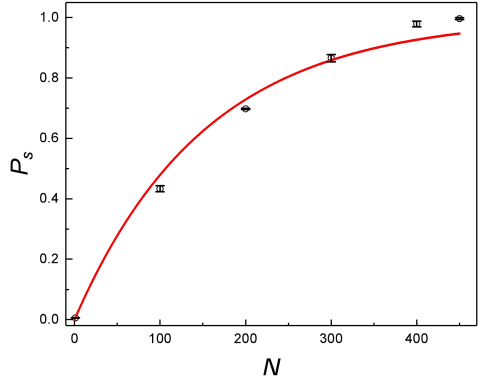

Fig. 5. Measurements of the feedback-based generation probability $P_S$ as a function of the write-sequence number $N$ for $\chi = 3\%$. The solid red curve is the fitting curve based on the function $P_S = \sum_{i=0}^{N-1} p_S (1-p_S)^i$ with experimental data $p_S = \chi \eta_S \approx 0.03 \times 0.2 = 0.006$. Error bars represent 1 standard deviation.

Using the feedback protocol with $N = 185$ write sequences ($T = 300\mu s$), we generate one SWE with a probability of ~66%. At a predetermined time of $T = 300\mu s$, i.e., $t_1 = 300\mu s$, we apply the first read pulse with an area of $\pi/2$ and then apply the second read pulse with an area of $\pi$ at $t_2$ [see Fig. 1(e)]. We then measure the anti-correlation parameter $\alpha$ between the two readouts as a function of the time separation ($\Delta t = t_2 - t_1$) and show the measurement results in Fig. 6. All the measured values are well below the classical bound of $\alpha \geq 1$, which implies that the quantum nature of the SW generated by the feed-back protocol may be preserved for more than 1 millisecond.

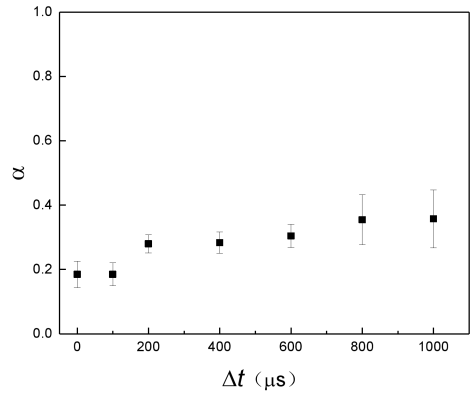

Fig. 6. The anti-correlation parameter $\alpha$ as a function of the time separation $\Delta t = t_2 - t_1$ for the case that the SWE is generated via the feed-back protocol with $N = 185$ write sequences. Error bars represent 1 standard deviation.

We then measure the anti-correlation parameter $\alpha$ between two readouts of a single SWE as a function of the write-sequence number $N$ for $\chi = 3\%$, the result is shown in Fig. 7. In the measurements, the two readouts are achieved by applying two read pulse, where, the first one has an area of $\pi/2$ and is applied at time $t_1 = T = N\delta t_w$ and the second one has an area of $\pi$ and is applied at time $t_2 = T + 1\mu s$ [see Fig. 1(e)]. The separation time between the two readouts is $\Delta t = t_2 - t_1 = 1\mu s$. In Fig. 7, $N = 450$ corresponds to the case that the SWE is generated with the probability of $0.996 \pm 0.003$ at predetermined time of $T = 730\mu s$. At this point, the measured autocorrelation $\alpha$ value is $0.35 \pm 0.09$. Such a measured $\alpha$ value is significantly below the classical bound, showing single-excitation character of the spin wave.

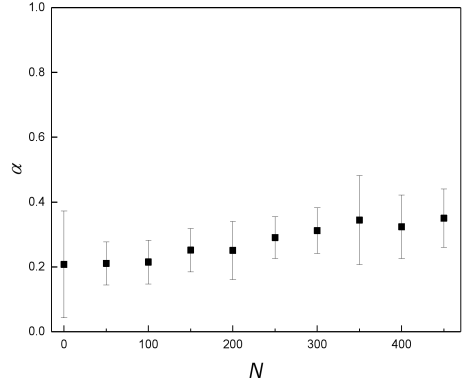

Fig. 7. The anti-correlation parameter $\alpha$ between two readouts of a single SWE as a function of the write-sequence number $N$ for $\chi = 3\%$. Error bars represent 1 standard deviation.

## 4. Conclusion

Based on the millisecond quantum memory and the feedback scheme, we generate one SWE with a near-unity ($0.996 \pm 0.003$) probability during a given time of $T = 730\mu s$. By applying a read pulse with power less than fully-retrieval value, we partially read out the SWE. Then the remained SW is fully read out by another read pulse after a storage time. The anti-correlation function between two readouts have been measured, which shows that the partial-retrieval operation on the SW is in the quantum regime. In the presented experiment, the required time for deterministically generating one SWE is several hundreds of microsecond, which is too long and will lead to a low quantum repeater rate. However, this is not a fundamental question. By decreasing the durations of write and cleaning pulses as well as the feedback delay time, the write-sequence period will be significantly shorten and then the required time will be greatly improved. In the presented experimental set up, some leakages of the laser beams which result from the imperfect switching off of the beams will lead to the light scattering in the atoms and then result in decoherence of the SWE. By improving the switching off, we may expect a further increase in the lifetime. The presented results represent a first step towards the realization of the improved DLCZ quantum repeater protocol [15].

**Funding**

Key Project of the Ministry of Science and Technology of China (Grant No. 2016YFA0301402); The National Natural Science Foundation of China (Grants: No. 11475109, No. 11604191), Shanxi"1331 Project" Key Subjects Construction and Program for Sanjin Scholars of Shanxi Province

**Disclosures**

The authors declare that there are no conflicts of interest related to this article.

**References**

1. H.-J. Briegel, W. Dür, J. I. Cirac, and P. Zoller, "Quantum Repeaters: The Role of Imperfect Local Operations in Quantum Communication," Phys. Rev. Lett. **81**, 5932−5935 (1998).
2. H. J. Kimble, "The quantum internet," Nature **453**, 1023−1030 (2008).
3. N. Sangouard, C. Simon, H. de Riedmatten, and N. Gisin, "Quantum repeaters based on atomic ensembles and linear optics," Rev. Mod. Phys. **83**, 33−80 (2011).
4. L.-M. Duan, M. Lukin, J. I. Cirac, and P. Zoller, "Long-distance quantum communication with atomic ensembles and linear optics," Nature **414**, 413−418 (2001)
5. C. W. Chou, H. de Riedmatten, D. Felinto, S. V. Polyakov, S. J. van Enk, and H. J. Kimble, "Measurement-induced entanglement for excitation stored in remote atomic ensembles," Nature **438**, 828−832 (2005).
6. J. Simon, H. Tanji, J. K. Thompson, and V. Vuletić, "Interfacing collective atomic excitations and single photons," Phys. Rev. Lett. **98**, 183601 (2007).
7. B. Zhao, Y.-A. Chen, X.-H. Bao, T. Strassel, C.-S. Chuu, X.-M. Jin, J. Schmiedmayer, Z.-S. Yuan, S. Chen, and J.-W. Pan, "A millisecond quantum memory for scalable quantum networks," Nat. Phys. **5**, 95−99 (2008).
8. R. Zhao, Y. O. Dudin, S. D. Jenkins, C. J. Campbell, D. N. Matsukevich, T. A. B. Kennedy, and A. Kuzmich, "Long-lived quantum memory," Nat. Phys. **5**, 100−104 (2008).
9. A. G. Radnaev, Y. O. Dudin, R. Zhao, H. H. Jen, S. D. Jenkins, A. Kuzmich, and T. A. B. Kennedy, "A quantum memory with telecom-wavelength conversion," Nat. Phys. **6**, 894−899 (2010).
10. X.-H. Bao, A. Reingruber, P. Dietrich, J. Rui, A. Dück, T. Strassel, L. Li, N.-L. Liu, B. Zhao, and J.-W. Pan, "Efficient and long-lived quantum memory with cold atoms inside a ring cavity," Nat. Phys. **8**, 517−521 (2012).
11. S.-J. Yang, X.-J. Wang, X.-H. Bao, and J.-W. Pan, "An efficient quantum light–matter interface with sub-second lifetime," Nat. Photonics **10**, 381−384 (2016).
12. C. W. Chou, S. V. Polyakov, A. Kuzmich, and H. J. Kimble, "Single-photon generation from stored excitation in an atomic ensemble," Phys. Rev. Lett. **92**, 213601 (2004).
13. D. N. Matsukevich, T. Chanelière, S. D. Jenkins, S.-Y. Lan, T. A. Kennedy, and A. Kuzmich, "Deterministic single photons via conditional quantum evolution," Phys. Rev. Lett. **97**, 013601 (2006).
14. S. Chen, Y.-A. Chen, T. Strassel, Z.-S. Yuan, B. Zhao, J. Schmiedmayer, and J.-W. Pan, "Deterministic and storable single-photon source based on a quantum memory," Phys. Rev. Lett. **97**, 173004 (2006).

15. N. Sangouard, C. Simon, B. Zhao, Y.-A. Chen, H. de Riedmatten, J.-W. Pan, and N. Gisin, "Robust and efficient quantum repeaters with atomic ensembles and linear optics," Phys. Rev. A **77**, 062301 (2008).
16. N. Sangouard, C. Simon, J. Minář, H. Zbinden, H. de Riedmatten, and N. Gisin, "Long-distance entanglement distribution with single-photon sources," Phys. Rev. A **76**, 050301 (2007).
17. Z.-B. Chen, B. Zhao, Y.-A. Chen, J. Schmiedmayer, and J.-W. Pan, "Fault-tolerant quantum repeater with atomic ensembles and linear optics," Phys. Rev. A **76**, 022329 (2007).
18. L. Jiang, J. M. Taylor, and M. D. Lukin, "Fast and robust approach to long-distance quantum communication with atomic ensembles," Phys. Rev. A **76**, 012301 (2007).
19. J. Minář, H. de Riedmatten, and N. Sangouard, "Quantum repeaters based on heralded qubit amplifiers," Phys. Rev. A **85**, 032313 (2012).
20. J. Li, M.-T. Zhou, B. Jing, X.-J. Wang, S.-J. Yang, X. Jiang, K. Mølmer, X.-H. Bao, and J.-W. Pan, "Hong-Ou-Mandel Interference between Two Deterministic Collective Excitations in an Atomic Ensemble" Phys. Rev. Lett. **117**, 180501 (2016).
21. L. Li, Y. O. Dudin, and A. Kuzmich, "Entanglement between light and an optical atomic excitation" Nature **498**, 466–469 (2013).
22. P. Farrera, G. Heinze, B. Albrecht, M. Ho, M. Chávez, C. Teo, N. Sangouard, and H. de Riedmatten, "Generation of single photons with highly tunable wave shape from a cold atomic ensemble" Nat. Commun. **7**, 13556 (2016).
23. Z. Xu, Y. Wu, L. Tian, L. Chen, Z. Zhang, Z. Yan, S. Li, H. Wang, C. Xie, and K. Peng, "Long lifetime and high-fidelity quantum memory of photonic polarization qubit by lifting zeeman degeneracy," Phys. Rev. Lett. **111**, 240503 (2013).